\pdfoutput=1
\documentclass[3p,a4paper,pdftex,twocolumn]{elsarticle}
\usepackage{amsmath,amssymb}
\usepackage{appendix}
\usepackage[T1]{fontenc}
\usepackage{graphics}
\usepackage[utf8]{inputenc}
\usepackage{microtype}
\usepackage{multirow}
\usepackage[nooneline]{subfigure}
\usepackage{textcomp}
\usepackage[textsize=footnotesize,textwidth=15mm]{todonotes}
\usepackage{xspace}
\usepackage{hyperref}

\biboptions{numbers, sort&compress}
\journal{Chemical Physics Letters}

\newlength{\figwidth}
\newcommand{\cis}{\textit{cis}\xspace}%

\newcommand{\eg}{e.\,g.}%
\newcommand{\FP}{3FP\xspace}%
\newcommand{\ie}{i.\,e.}%

\newcommand{\mueff}{\ensuremath{\mu_{\text{eff}}}\xspace}
\newcommand{\invcm}{\ensuremath{\text{cm}^{-1}}\xspace}

\newcommand{\trans}{\textit{trans}\xspace}%

\begin{document}
\begin{frontmatter}
   \title{Spatially separated polar samples of the \\ \cis and \trans conformers of 3-fluorophenol}%
   \author[cfel,cui]{Thomas Kierspel}%
   \author[cfel]{Daniel A.\ Horke}%
   \author[cfel]{Yuan-Pin Chang}%
   \author[cfel,uhh,cui]{Jochen Küpper\corref{cor}}%
   \cortext[cor]{Corresponding author}\ead{jochen.kuepper@cfel.de}\ead[url]{http://desy.cfel.de/cid/cmi}%
   \address[cfel]{Center for Free-Electron Laser Science, DESY, Notkestrasse 85, 22607 Hamburg, Germany}%
   \address[uhh]{\mbox{Department of Physics, University of Hamburg, Luruper Chausse 149, 22761 Hamburg, Germany}}%
   \address[cui]{\mbox{The Hamburg Center for Ultrafast Imaging, Luruper Chaussee 149, 22761 Hamburg, Germany}}%
   \begin{abstract}\noindent
      We demonstrate the spatial separation of the \cis- and \trans-conformers of 3-fluorophenol in
      the gas phase based on their distinct electric dipole moments. For both conformers we create
      very polar samples of their lowest-energy rotational quantum states. A >95~\% pure beam of
      \trans-3-fluorophenol and a >90~\% pure beam of the lowest-energy rotational states of the
      less polar \cis-3-fluorophenol were obtained for helium and neon supersonic expansions,
      respectively. This is the first demonstration of the spatial separation of the lowest-energy
      rotational states of the least polar conformer, which is necessary for strong alignment and
      orientation of all individual conformers.
   \end{abstract}
\end{frontmatter}

\section{Introduction}
\label{sec:introduction}
Complex molecules often exhibit different structural isomers (conformers), even under the conditions
of a cold molecular beam~\cite{Rizzo:JCP83:4819}. For a variety of upcoming novel experiments like
photoelectron~\cite{Krasniqi:PRA81:033411, Boll:PRL:2013}, electron~\cite{Sciaini:RPP74:096101,
   Hensley:PRL109:133202}, and x-ray diffraction~\cite{Barty:ARPC64:415, Kuepper:LCLSdibn:inprep}
imaging of gas-phase molecules, for conformer-specific chemical reaction
studies~\cite{Chang:Science342:2013}, or for mixed-field orientation
experiments~\cite{Stapelfeldt:RMP75:543, Holmegaard:PRL102:023001, Ghafur:NatPhys5:289}, pure
species-selected molecular samples with all molecules in the lowest-energy rotational states are
highly advantageous or simply necessary. Using strong inhomogeneous electric fields it is possible
to spatially separate individual conformers~\cite{Filsinger:PRL100:133003, Filsinger:ACIE48:6900},
specific molecular clusters~\cite{Trippel:PRA86:033202}, as well as the individual quantum states of
neutral polar molecules~\cite{Reuss:StateSelection, Filsinger:JCP131:064309, Putzke:PCCP13:18962}.
While the electric deflector was exploited for the spatial separation of individual
conformers~\cite{Filsinger:ACIE48:6900} and for the generation of highly polar samples of
single-conformer molecules~\cite{Holmegaard:PRL102:023001}, it was not \emph{a priory} clear whether
it would be possible to create pure samples of the most polar low-rotational-energy states of any
conformer but the most polar one.

\begin{figure}[t]
   \centering
   \includegraphics[width=\linewidth]{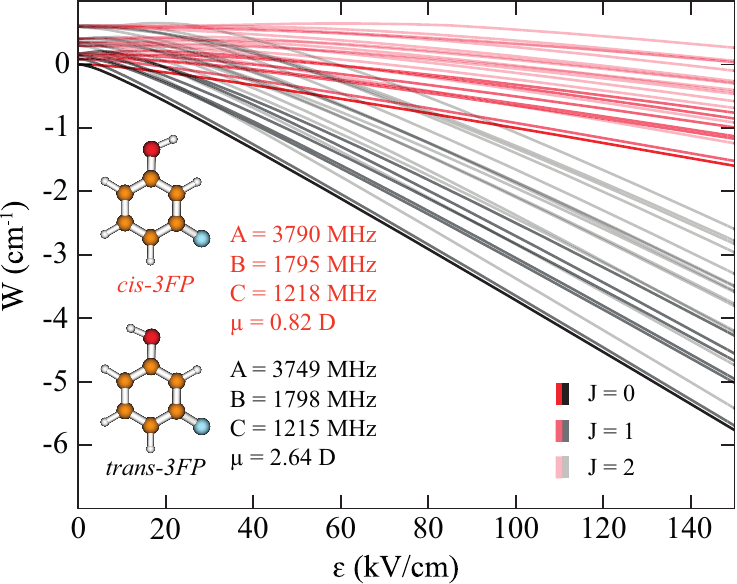}
   \caption{Stark-energies for the lowest rovibronic quantum states ($J<3$) of the \cis- (red) and
      \trans-conformer (black) of \FP as a function of the electric field strength. The inset shows
      the molecular structures, the rotational constants \cite{Jaman:JMolSpec86:269,
         Dutta:Pramana24:499}, and the calculated dipole moments of these conformers.}
   \label{fig:Stark}
\end{figure}
Here, we demonstrate the generation of conformer-selected and very polar ensembles of both
conformers of 3-fluorophenol ($\mathrm{C_6H_5OF}$, \FP). \FP is a prototypical large molecule with
two stable conformers that differ by their orientation of the OH functional group.
Conformer-selected and three-dimensionally oriented~\cite{Nevo:PCCP11:9912} ensembles of \FP are
good candidates for the investigation of conformer-interconversion reactions in an ultrafast laser
pump, x-ray probe experiment, in which structural information would be accessible through
photoelectron holography following F(1s) ionization~\cite{Krasniqi:PRA81:033411, Boll:PRL:2013}.

The structures of \cis and \trans \FP are shown in \autoref{fig:Stark}, together with calculated
energies of their lowest rotational states in an electric field~\cite{Chang:CPC:2013}. At relevant
electric field strengths, all states are high-field seeking.\footnote{Molecules in high-field
   seeking states have a negative Stark energy shift, \ie, their potential energy decreases with
   electric field strength. Thus, they have positive effective dipole moments (\emph{vide infra}).
   Therefore, they orient along the electric field and they are attracted to regions of stronger
   electric field in order to minimize their energy.} Due to the different electric dipole moments
of $\mu=0.82$~D and 2.64~D for \cis and \trans, respectively, the energy dependence of the molecular
eigenstates as a function of electric field strength is quite different for the two conformers. The
projection of the molecule-fixed dipole moment onto the electric field axis is the effective dipole
moment $\mueff=-\partial{W}/\partial{\epsilon}$. The force exerted by an electric field on a polar
molecule is $\vec{F}=- \mueff(\epsilon)\cdot\vec{\nabla}\epsilon$. Thus, for a given inhomogeneous
electric field \trans-\FP is deflected more than \cis-\FP. Moreover, the lowest energy rotational
states are generally more polar, \ie, have a larger effective dipole moment \mueff, than
higher-energy rotational states~\cite{Holmegaard:PRL102:023001} and, therefore, are deflected more.

\section{Experimental details}
\label{sec:experiment}
\begin{figure}[t]
   \centering
   \includegraphics[width=\linewidth]{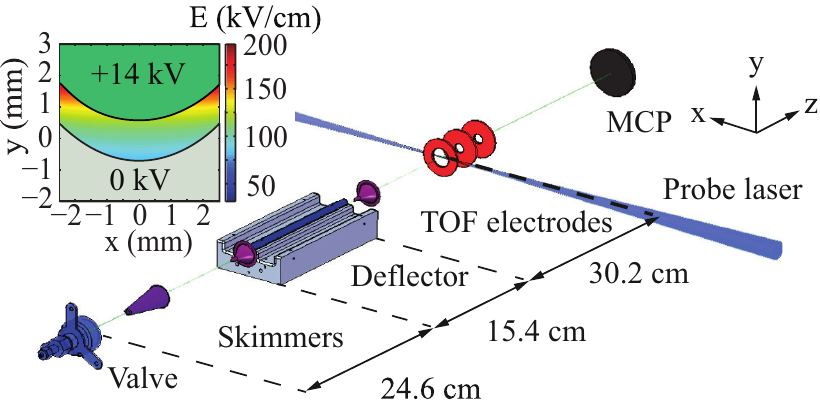}
   \caption{Schematic of the experimental setup consisting of a pulsed molecular beam source, a
      deflector, and a time-of-flight mass spectrometer. The inset shows a cut through the deflector
      and depicts the generated electric field.}
   \label{fig:expSetup}
\end{figure}
\autoref{fig:expSetup} shows a schematic of the experimental setup~\cite{Filsinger:JCP131:064309}.
In brief, \FP was placed in a room-temperature Even-Lavie valve~\cite{Even:JCP112:8068}. The
molecules are coexpanded into the vacuum chamber at a repetition rate of 20~Hz in 50~bar of helium
or 25~bar of neon, resulting in a supersonic expansion with a rotational temperature of the
molecules of 1.5~K and 1~K, respectively. The molecular beam was collimated by two skimmers with
diameters of 2~mm and 1~mm that were placed 5.5~cm and 21.5~cm downstream of the valve. Behind the
second skimmer the molecular beam entered the electric deflector that provided an inhomogeneous
electric field to disperse the molecular beam. A third skimmer with 1.5~mm diameter was placed
directly behind the deflector. A time-of-flight mass-spectrometer was placed 30~cm downstream the
deflector. A tunable pulsed dye laser (Fine Adjustment Pulsare Pro) provided approximately
400~\textmu{J} per pulse in a soft focus at the electronic origin transitions of 36623~\invcm and
36830~\invcm for \cis and \trans, respectively~\cite{Fujimaki:JCP110:4238}. Individual conformers
were selectively detected through resonance-enhanced two-photon ionization (R2PI) mass spectrometry.
For the measurements of spatial molecular beam profiles the laser is sampled by moving the focusing
lens in steps of 200~\textmu{m} in the y-direction.

\emph{Ab initio} calculations (B3LYP/aug-cc-pVTZ) predict the \trans conformer to be about
55~cm$^{-1}$ more stable than the \cis conformer, in agreement with the 85~cm$^{-1}$ energy
difference derived from a one-dimensionally torsional potential based on far-infrared
spectroscopy~\cite{Manocha:JPC77:2094}. The corresponding expected relative abundances of the \trans
and \cis conformers in the molecular beam of approximately 2:1 agree with the experimentally
obtained R2PI signal levels.

\section{Results and discussions}
\begin{figure}[t]
   \centering
   \includegraphics[width=\linewidth]{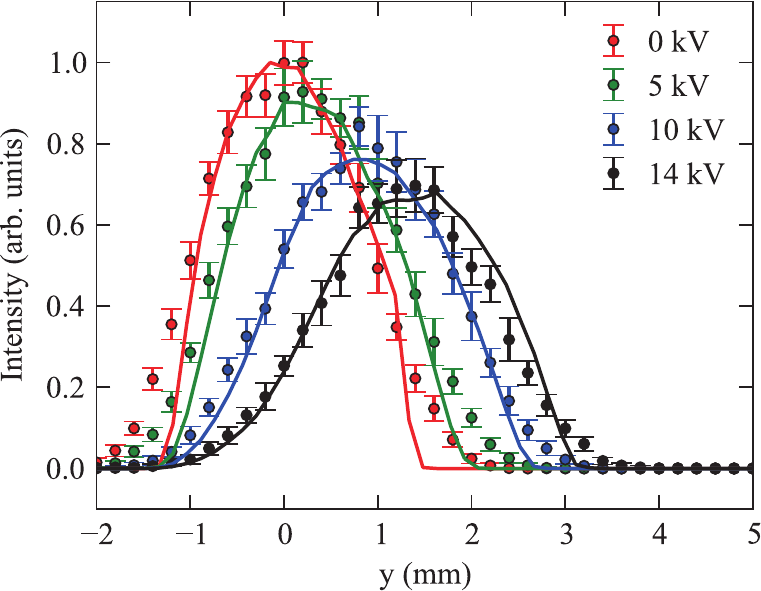}
   \caption{Spatial profiles for \trans-\FP seeded in helium for different voltages applied to the
      deflector. The solid lines indicate simulated deflection profiles, the symbols are
      experimental values and the depicted ranges are their $1\sigma$ standard deviation.}
   \label{fig:HeDef}
\end{figure}
\autoref{fig:HeDef} shows the vertical molecular beam profile for \trans-\FP seeded in helium for
different voltages applied to the deflector. The height of the undeflected molecular beam (0~kV) is
2~mm, defined by the mechanical apertures of the deflector and the skimmers. When a voltage of 5~kV
is applied to the deflector the beam is deflected upwards by approximately 0.4~mm for the
\trans-conformer. Increasing the voltage to 10~kV and 14~kV, the deflection increases to 0.8~mm and
1.4~mm, respectively. Solid lines show simulated spatial profiles obtained using the libcolmol
package~\cite{Filsinger:JCP131:064309}. Only a single free parameter, the peak intensity of the
undeflected beam, was adjusted to the experiment.

\begin{figure}[t]
   \centering
   \includegraphics[width=\linewidth]{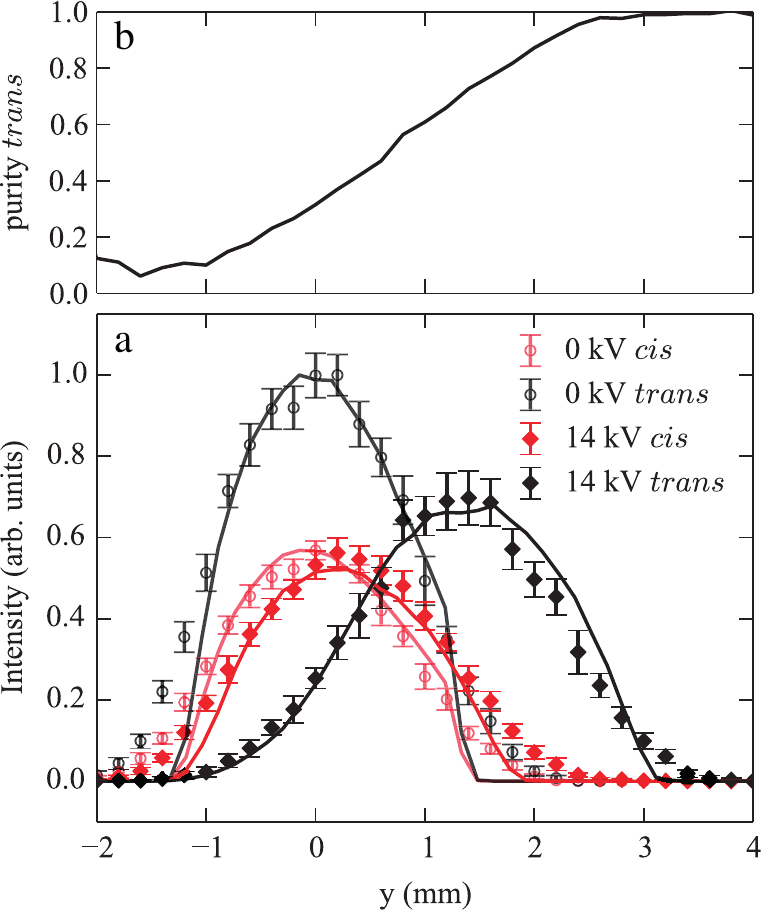}
   \caption{a) Field-free (0~kV) and deflected (14~kV) spatial beam profiles for the \cis and the
      \trans conformers of \FP seeded in helium. b) The fractional population of the \trans isomer
      in the beam.}
   \label{fig:HeDef1}
\end{figure}
\autoref{fig:HeDef1}\,a shows the deflection profiles of both conformers of \FP seeded in helium for
applied voltages of 0~V and 14~kV to the deflector. The \cis-conformer was deflected significantly
less than the \trans-conformer. This leads to a spatial separation of the \trans-conformer from the
\cis-conformer and from the original beam, \eg, the seed gas. The purity of the \trans sample at
various heights is shown in \autoref{fig:HeDef1}\,b. Assuming similar excitation and ionization
cross-sections for the two conformers, a beam of \trans-\FP with a purity >95~$\%$ was obtained in
the range of 2.5~mm to 3.5~mm.

\begin{figure}[t]
   \centering
   \includegraphics[width=\linewidth]{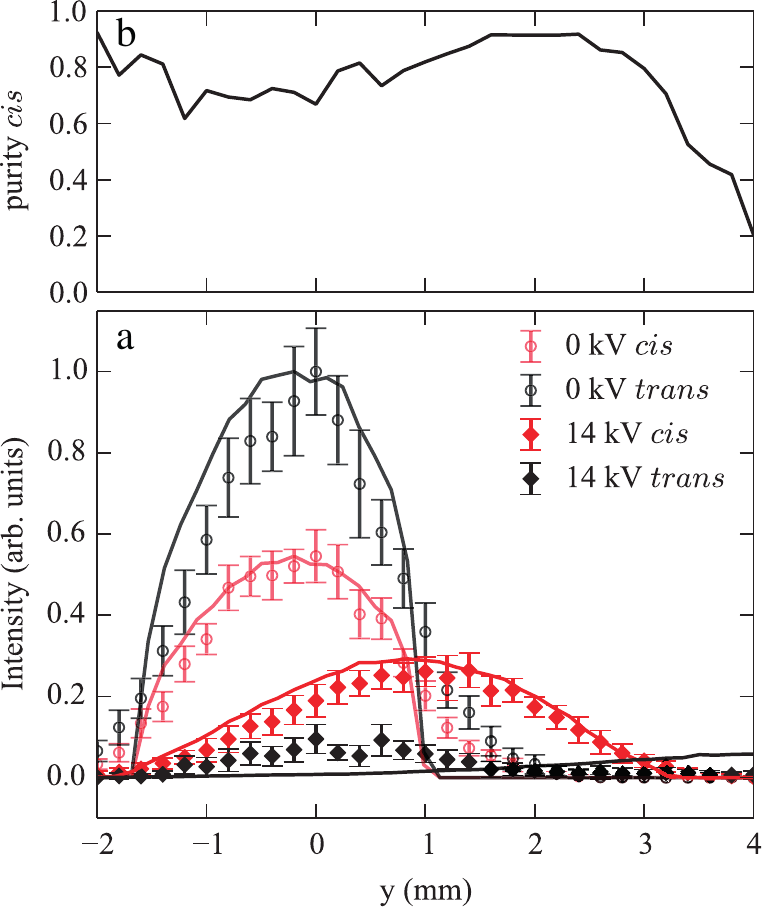}
   \caption{a) Field-free (0~kV) and deflected (14~kV) spatial beam profiles for the \cis and the
      \trans conformers of \FP seeded in neon. b) The fractional population of the \cis isomer in
      the beam.}
   \label{fig:NeDef}
\end{figure}

Around $y=-0.8$~mm a nearly pure beam of \cis-\FP was obtained, similar to previous
measurements~\cite{Filsinger:ACIE48:6900}. However, these samples correspond to high-energy
rotational states that are not suited for the envisioned novel orientation and imaging experiments.
Moreover, these molecules were still immersed in the atomic seed gas. In order to create a pure beam
of low-energy rotational states of \cis-\FP we increased the deflection by seeding the molecules in
neon, resulting in a molecular beam with a speed of 900~m/s, half the speed of the helium expansion.
The doubled interaction time resulted in a stronger deflection, as shown in \autoref{fig:NeDef}\,a.
For an applied voltage of 14~kV both conformers were strongly deflected. For the originally very
cold molecular beam the deflection of \trans-\FP was so strong that most molecules crash into the
rod electrode and are lost from the beam. For the \cis conformer, however, a strong deflection out
of the seed gas beam was observed. For $y=1.5$--$2.5$~mm a beam of \cis-\FP with a purity >90~\%
with all molecules in the lowest-energy rotational states was obtained, see \autoref{fig:NeDef}\,b,
even though this conformer was less abundant in the original beam (\emph{vide supra}). The remaining
signal strength at position 1.5~mm was 40~$\%$ of the maximum of the field-free case. The amplitudes
for each conformer are normalized to the laser intensity. The remaining contribution from \trans-\FP
for $y<2$~mm is attributed to warm molecules that diffuse into the detection region after colliding
with the deflector rod.

\section{Conclusions}
\label{sec:conclusions}
In summary, we have demonstrated the spatial separation of the \cis and \trans conformers of \FP and
the generation of cold, low-energy rotational state samples of both conformers of \FP. These results
demonstrate the feasibility to create pure samples of both conformers of this prototypical large
molecule using the electric deflector, allowing for strong three-dimensional alignment and
mixed-field orientation~\cite{Nevo:PCCP11:9912}. These samples will allow direct molecular-frame
imaging experiments on all structural isomers to disentangle the structure-function relationship for
the conformer interconversion. This could be investigated through molecular frame photoelectron
angular distributions and electron or x-ray diffraction experiments~\cite{Holmegaard:NatPhys6:428,
   Boll:PRL:2013, Hensley:PRL109:133202, Filsinger:PCCP13:2076, Barty:ARPC64:415,
   Kuepper:LCLSdibn:inprep}.

\section*{Acknowledgments}
We thank Songhee Han for initial \emph{ab initio} calculation of the molecular constants of the two
conformers. This work has been supported by the excellence cluster ``The Hamburg Center for
Ultrafast Imaging -- Structure, Dynamics and Control of Matter at the Atomic Scale'' of the Deutsche
Forschungsgemeinschaft and the Helmholtz Virtual Institute ``Dynamic Pathways in Multidimensional
Landscapes''.

\bibliographystyle{model1-num-names}
\bibliography{string,cmi}

\begin{thebibliography}{26}
\expandafter\ifx\csname natexlab\endcsname\relax\def\natexlab#1{#1}\fi
\providecommand{\url}[1]{\texttt{#1}}
\providecommand{\href}[2]{#2}
\providecommand{\path}[1]{#1}
\providecommand{\DOIprefix}{doi:}
\providecommand{\ArXivprefix}{arXiv:}
\providecommand{\URLprefix}{URL: }
\providecommand{\Pubmedprefix}{pmid:}
\providecommand{\doi}[1]{\href{http://dx.doi.org/#1}{\path{#1}}}
\providecommand{\Pubmed}[1]{\href{pmid:#1}{\path{#1}}}
\providecommand{\bibinfo}[2]{#2}
\ifx\xfnm\relax \def\xfnm[#1]{\unskip,\space#1}\fi
\bibitem[{Rizzo et~al.(1985)Rizzo, Park, Peteanu, and Levy}]{Rizzo:JCP83:4819}
\bibinfo{author}{T.~R. Rizzo}, \bibinfo{author}{Y.~D. Park},
  \bibinfo{author}{L.~Peteanu}, \bibinfo{author}{D.~H. Levy},
\newblock \bibinfo{title}{Electronic spectrum of the amino acid tryptophan
  cooled in a supersonic molecular beam},
\newblock \bibinfo{journal}{J.\ Chem.\ Phys.} \bibinfo{volume}{83}
  (\bibinfo{year}{1985}) \bibinfo{pages}{4819--4820}.
\bibitem[{Krasniqi et~al.(2010)Krasniqi, Najjari, Str\"{u}der, Rolles, Voitkiv,
  and Ullrich}]{Krasniqi:PRA81:033411}
\bibinfo{author}{F.~Krasniqi}, \bibinfo{author}{B.~Najjari},
  \bibinfo{author}{L.~Str\"{u}der}, \bibinfo{author}{D.~Rolles},
  \bibinfo{author}{A.~Voitkiv}, \bibinfo{author}{J.~Ullrich},
\newblock \bibinfo{title}{Imaging molecules from within: {U}ltrafast
  angstr\"{o}m-scale structure determination of molecules via photoelectron
  holography using free-electron lasers},
\newblock \bibinfo{journal}{Phys.\ Rev.\ A} \bibinfo{volume}{81}
  (\bibinfo{year}{2010}) \bibinfo{pages}{033411}.
\bibitem[{Boll et~al.(2013)Boll, Anielski, Bostedt, Bozek, Christensen, Coffee,
  De, Decleva, Epp, Erk, Foucar, Krasniqi, K\"upper, Rouz\'{e}e, Rudek,
  Rudenko, Schorb, Stapelfeldt, Stener, Stern, Techert, Trippel, Vrakking,
  Ullrich, and Rolles}]{Boll:PRL:2013}
\bibinfo{author}{R.~Boll}, \bibinfo{author}{D.~Anielski},
  \bibinfo{author}{C.~Bostedt}, \bibinfo{author}{J.~D. Bozek},
  \bibinfo{author}{L.~Christensen}, \bibinfo{author}{R.~Coffee},
  \bibinfo{author}{S.~De}, \bibinfo{author}{P.~Decleva}, \bibinfo{author}{S.~W.
  Epp}, \bibinfo{author}{B.~Erk}, \bibinfo{author}{L.~Foucar},
  \bibinfo{author}{F.~Krasniqi}, \bibinfo{author}{J.~K\"upper},
  \bibinfo{author}{A.~Rouz\'{e}e}, \bibinfo{author}{B.~Rudek},
  \bibinfo{author}{A.~Rudenko}, \bibinfo{author}{S.~Schorb},
  \bibinfo{author}{H.~Stapelfeldt}, \bibinfo{author}{M.~Stener},
  \bibinfo{author}{S.~Stern}, \bibinfo{author}{S.~Techert},
  \bibinfo{author}{S.~Trippel}, \bibinfo{author}{M.~J.~J. Vrakking},
  \bibinfo{author}{J.~Ullrich}, \bibinfo{author}{D.~Rolles},
\newblock \bibinfo{title}{Femtosecond photoelectron diffraction on
  laser-aligned molecules: Freeze frames of a molecular movie}
  (\bibinfo{year}{2013}). \bibinfo{note}{Submitted}.
\bibitem[{Sciaini and Miller(2011)}]{Sciaini:RPP74:096101}
\bibinfo{author}{G.~Sciaini}, \bibinfo{author}{R.~J.~D. Miller},
\newblock \bibinfo{title}{Femtosecond electron diffraction: heralding the era
  of atomically resolved dynamics},
\newblock \bibinfo{journal}{Rep.\ Prog.\ Phys.} \bibinfo{volume}{74}
  (\bibinfo{year}{2011}) \bibinfo{pages}{096101}.
\bibitem[{Hensley et~al.(2012)Hensley, Yang, and
  Centurion}]{Hensley:PRL109:133202}
\bibinfo{author}{C.~J. Hensley}, \bibinfo{author}{J.~Yang},
  \bibinfo{author}{M.~Centurion},
\newblock \bibinfo{title}{Imaging of isolated molecules with ultrafast electron
  pulses},
\newblock \bibinfo{journal}{Phys.\ Rev.\ Lett.} \bibinfo{volume}{109}
  (\bibinfo{year}{2012}) \bibinfo{pages}{133202}.
\bibitem[{Barty et~al.(2013)Barty, K{\"u}pper, and Chapman}]{Barty:ARPC64:415}
\bibinfo{author}{A.~Barty}, \bibinfo{author}{J.~K{\"u}pper},
  \bibinfo{author}{H.~N. Chapman},
\newblock \bibinfo{title}{Molecular imaging using x-ray free-electron lasers},
\newblock \bibinfo{journal}{Annu.\ Rev.\ Phys.\ Chem.} \bibinfo{volume}{64}
  (\bibinfo{year}{2013}) \bibinfo{pages}{415--435}.
\bibitem[{K{\"u}pper et~al.(2013)K{\"u}pper, Stern, Holmegaard, Filsinger,
  Rouz\'{e}e, Rolles, Rudenko, Johnsson, Martin, Adolph, Aquila, Bajt, Barty,
  Bostedt, Bozek, Caleman, Coffee, Coppola, Delmas, Epp, Erk, Foucar,
  Gorkhover, Gumprecht, Hartmann, Hartmann, Hauser, Holl, H{\"o}mke, Kimmel,
  Krasniqi, K{\"u}hnel, Maurer, Messerschmidt, Moshammer, Reich, Rudek, Santra,
  Schlichting, Schmidt, Schorb, Schulz, Soltau, Spence, Starodub, Str{\"u}der,
  Th{\o}gersen, Vrakking, Weidenspointner, White, Wunderer, Meijer, Ullrich,
  Stapelfeldt, and Chapman}]{Kuepper:LCLSdibn:inprep}
\bibinfo{author}{J.~K{\"u}pper}, \bibinfo{author}{S.~Stern},
  \bibinfo{author}{L.~Holmegaard}, \bibinfo{author}{F.~Filsinger},
  \bibinfo{author}{A.~Rouz\'{e}e}, \bibinfo{author}{D.~Rolles},
  \bibinfo{author}{A.~Rudenko}, \bibinfo{author}{P.~Johnsson},
  \bibinfo{author}{A.~V. Martin}, \bibinfo{author}{M.~Adolph},
  \bibinfo{author}{A.~Aquila}, \bibinfo{author}{S.~Bajt},
  \bibinfo{author}{A.~Barty}, \bibinfo{author}{C.~Bostedt},
  \bibinfo{author}{J.~Bozek}, \bibinfo{author}{C.~Caleman},
  \bibinfo{author}{R.~Coffee}, \bibinfo{author}{N.~Coppola},
  \bibinfo{author}{T.~Delmas}, \bibinfo{author}{S.~Epp},
  \bibinfo{author}{B.~Erk}, \bibinfo{author}{L.~Foucar},
  \bibinfo{author}{T.~Gorkhover}, \bibinfo{author}{L.~Gumprecht},
  \bibinfo{author}{A.~Hartmann}, \bibinfo{author}{R.~Hartmann},
  \bibinfo{author}{G.~Hauser}, \bibinfo{author}{P.~Holl},
  \bibinfo{author}{A.~H{\"o}mke}, \bibinfo{author}{N.~Kimmel},
  \bibinfo{author}{F.~Krasniqi}, \bibinfo{author}{K.-U. K{\"u}hnel},
  \bibinfo{author}{J.~Maurer}, \bibinfo{author}{M.~Messerschmidt},
  \bibinfo{author}{R.~Moshammer}, \bibinfo{author}{C.~Reich},
  \bibinfo{author}{B.~Rudek}, \bibinfo{author}{R.~Santra},
  \bibinfo{author}{I.~Schlichting}, \bibinfo{author}{C.~Schmidt},
  \bibinfo{author}{S.~Schorb}, \bibinfo{author}{J.~Schulz},
  \bibinfo{author}{H.~Soltau}, \bibinfo{author}{J.~C.~H. Spence},
  \bibinfo{author}{D.~Starodub}, \bibinfo{author}{L.~Str{\"u}der},
  \bibinfo{author}{J.~Th{\o}gersen}, \bibinfo{author}{M.~J.~J. Vrakking},
  \bibinfo{author}{G.~Weidenspointner}, \bibinfo{author}{T.~A. White},
  \bibinfo{author}{C.~Wunderer}, \bibinfo{author}{G.~Meijer},
  \bibinfo{author}{J.~Ullrich}, \bibinfo{author}{H.~Stapelfeldt},
  \bibinfo{author}{H.~N. Chapman},
\newblock \bibinfo{title}{Coherent diffractive imaging of controlled ensembles
  of isolated gas-phase molecules}  (\bibinfo{year}{2013}).
  \bibinfo{note}{Submitted, arXiv:1307.4577 [physics]}.
\bibitem[{Chang et~al.(2013)Chang, D{\l}ugo{\l}{\k e}cki, K{\"u}pper, R\"osch,
  Wild, and Willitsch}]{Chang:Science342:2013}
\bibinfo{author}{Y.-P. Chang}, \bibinfo{author}{K.~D{\l}ugo{\l}{\k e}cki},
  \bibinfo{author}{J.~K{\"u}pper}, \bibinfo{author}{D.~R\"osch},
  \bibinfo{author}{D.~Wild}, \bibinfo{author}{S.~Willitsch},
\newblock \bibinfo{title}{Specific {C}hemical {R}eactivities of {S}patially
  {S}eparated 3-{A}minophenol {C}onformers with {C}old {C}a$^+$ {I}ons},
\newblock \bibinfo{journal}{Science} \bibinfo{volume}{342}
  (\bibinfo{year}{2013}) \bibinfo{pages}{98--101}.
\bibitem[{Stapelfeldt and Seideman(2003)}]{Stapelfeldt:RMP75:543}
\bibinfo{author}{H.~Stapelfeldt}, \bibinfo{author}{T.~Seideman},
\newblock \bibinfo{title}{Colloquium: Aligning molecules with strong laser
  pulses},
\newblock \bibinfo{journal}{Rev.\ Mod.\ Phys.} \bibinfo{volume}{75}
  (\bibinfo{year}{2003}) \bibinfo{pages}{543--557}.
\bibitem[{Holmegaard et~al.(2009)Holmegaard, Nielsen, Nevo, Stapelfeldt,
  Filsinger, K\"upper, and Meijer}]{Holmegaard:PRL102:023001}
\bibinfo{author}{L.~Holmegaard}, \bibinfo{author}{J.~H. Nielsen},
  \bibinfo{author}{I.~Nevo}, \bibinfo{author}{H.~Stapelfeldt},
  \bibinfo{author}{F.~Filsinger}, \bibinfo{author}{J.~K\"upper},
  \bibinfo{author}{G.~Meijer},
\newblock \bibinfo{title}{Laser-induced alignment and orientation of
  quantum-state-selected large molecules},
\newblock \bibinfo{journal}{Phys.\ Rev.\ Lett.} \bibinfo{volume}{102}
  (\bibinfo{year}{2009}) \bibinfo{pages}{023001}.
\bibitem[{Ghafur et~al.(2009)Ghafur, Rouzee, Gijsbertsen, Siu, Stolte, and
  Vrakking}]{Ghafur:NatPhys5:289}
\bibinfo{author}{O.~Ghafur}, \bibinfo{author}{A.~Rouzee},
  \bibinfo{author}{A.~Gijsbertsen}, \bibinfo{author}{W.~K. Siu},
  \bibinfo{author}{S.~Stolte}, \bibinfo{author}{M.~J.~J. Vrakking},
\newblock \bibinfo{title}{Impulsive orientation and alignment of
  quantum-state-selected {NO} molecules},
\newblock \bibinfo{journal}{Nat. Phys.} \bibinfo{volume}{5}
  (\bibinfo{year}{2009}) \bibinfo{pages}{289--293}.
\bibitem[{Filsinger et~al.(2008)Filsinger, Erlekam, von Helden, K\"upper, and
  Meijer}]{Filsinger:PRL100:133003}
\bibinfo{author}{F.~Filsinger}, \bibinfo{author}{U.~Erlekam},
  \bibinfo{author}{G.~von Helden}, \bibinfo{author}{J.~K\"upper},
  \bibinfo{author}{G.~Meijer},
\newblock \bibinfo{title}{Selector for structural isomers of neutral
  molecules},
\newblock \bibinfo{journal}{Phys.\ Rev.\ Lett.} \bibinfo{volume}{100}
  (\bibinfo{year}{2008}) \bibinfo{pages}{133003}.
\bibitem[{Filsinger et~al.(2009)Filsinger, K\"upper, Meijer, Hansen, Maurer,
  Nielsen, Holmegaard, and Stapelfeldt}]{Filsinger:ACIE48:6900}
\bibinfo{author}{F.~Filsinger}, \bibinfo{author}{J.~K\"upper},
  \bibinfo{author}{G.~Meijer}, \bibinfo{author}{J.~L. Hansen},
  \bibinfo{author}{J.~Maurer}, \bibinfo{author}{J.~H. Nielsen},
  \bibinfo{author}{L.~Holmegaard}, \bibinfo{author}{H.~Stapelfeldt},
\newblock \bibinfo{title}{Pure samples of individual conformers: the separation
  of stereo-isomers of complex molecules using electric fields},
\newblock \bibinfo{journal}{Angew.\ Chem.\ Int.\ Ed.} \bibinfo{volume}{48}
  (\bibinfo{year}{2009}) \bibinfo{pages}{6900--6902}.
\bibitem[{Trippel et~al.(2012)Trippel, Chang, Stern, Mullins, Holmegaard, and
  K{\"u}pper}]{Trippel:PRA86:033202}
\bibinfo{author}{S.~Trippel}, \bibinfo{author}{Y.-P. Chang},
  \bibinfo{author}{S.~Stern}, \bibinfo{author}{T.~Mullins},
  \bibinfo{author}{L.~Holmegaard}, \bibinfo{author}{J.~K{\"u}pper},
\newblock \bibinfo{title}{Spatial separation of state- and size-selected
  neutral clusters},
\newblock \bibinfo{journal}{Phys.\ Rev.\ A} \bibinfo{volume}{86}
  (\bibinfo{year}{2012}) \bibinfo{pages}{033202}.
\bibitem[{Reuss(1988)}]{Reuss:StateSelection}
\bibinfo{author}{J.~Reuss},
\newblock \bibinfo{title}{{S}tate {S}election by {N}onoptical {M}ethods},
\newblock in: \bibinfo{editor}{G.~Scoles} (Ed.), \bibinfo{booktitle}{Atomic and
  molecular beam methods}, volume~\bibinfo{volume}{1},
  \bibinfo{publisher}{Oxford University Press}, \bibinfo{address}{New York, NY,
  USA}, \bibinfo{year}{1988}, pp. \bibinfo{pages}{276--292}.
\bibitem[{Filsinger et~al.(2009)Filsinger, K\"upper, Meijer, Holmegaard,
  Nielsen, Nevo, Hansen, and Stapelfeldt}]{Filsinger:JCP131:064309}
\bibinfo{author}{F.~Filsinger}, \bibinfo{author}{J.~K\"upper},
  \bibinfo{author}{G.~Meijer}, \bibinfo{author}{L.~Holmegaard},
  \bibinfo{author}{J.~H. Nielsen}, \bibinfo{author}{I.~Nevo},
  \bibinfo{author}{J.~L. Hansen}, \bibinfo{author}{H.~Stapelfeldt},
\newblock \bibinfo{title}{Quantum-state selection, alignment, and orientation
  of large molecules using static electric and laser fields},
\newblock \bibinfo{journal}{J.\ Chem.\ Phys.} \bibinfo{volume}{131}
  (\bibinfo{year}{2009}) \bibinfo{pages}{064309}.
\bibitem[{Putzke et~al.(2011)Putzke, Filsinger, Haak, K\"upper, and
  Meijer}]{Putzke:PCCP13:18962}
\bibinfo{author}{S.~Putzke}, \bibinfo{author}{F.~Filsinger},
  \bibinfo{author}{H.~Haak}, \bibinfo{author}{J.~K\"upper},
  \bibinfo{author}{G.~Meijer},
\newblock \bibinfo{title}{Rotational-state-specific guiding of large
  molecules},
\newblock \bibinfo{journal}{Phys.\ Chem.\ Chem.\ Phys.} \bibinfo{volume}{13}
  (\bibinfo{year}{2011}) \bibinfo{pages}{18962}.
\bibitem[{Jaman et~al.(1981)Jaman, Nandi, and Ghosh}]{Jaman:JMolSpec86:269}
\bibinfo{author}{A.~I. Jaman}, \bibinfo{author}{R.~N. Nandi},
  \bibinfo{author}{D.~K. Ghosh},
\newblock \bibinfo{title}{Microwave spectrum of 3-fluorophenol},
\newblock \bibinfo{journal}{J.\ Mol.\ Spec.} \bibinfo{volume}{86}
  (\bibinfo{year}{1981}) \bibinfo{pages}{269--274}.
\bibitem[{Dutta and Jaman(1985)}]{Dutta:Pramana24:499}
\bibinfo{author}{A.~Dutta}, \bibinfo{author}{A.~I. Jaman},
\newblock \bibinfo{title}{Microwave spectrum of cis 3-fluorophenol},
\newblock \bibinfo{journal}{Pramana -- J. Phys.} \bibinfo{volume}{24}
  (\bibinfo{year}{1985}) \bibinfo{pages}{499--502}.
\bibitem[{Nevo et~al.(2009)Nevo, Holmegaard, Nielsen, Hansen, Stapelfeldt,
  Filsinger, Meijer, and K\"upper}]{Nevo:PCCP11:9912}
\bibinfo{author}{I.~Nevo}, \bibinfo{author}{L.~Holmegaard},
  \bibinfo{author}{J.~H. Nielsen}, \bibinfo{author}{J.~L. Hansen},
  \bibinfo{author}{H.~Stapelfeldt}, \bibinfo{author}{F.~Filsinger},
  \bibinfo{author}{G.~Meijer}, \bibinfo{author}{J.~K\"upper},
\newblock \bibinfo{title}{Laser-induced 3{D} alignment and orientation of
  quantum state-selected molecules},
\newblock \bibinfo{journal}{Phys.\ Chem.\ Chem.\ Phys.} \bibinfo{volume}{11}
  (\bibinfo{year}{2009}) \bibinfo{pages}{9912--9918}.
\bibitem[{Chang et~al.(2013)Chang, Filsinger, Sartakov, and
  Küpper}]{Chang:CPC:2013}
\bibinfo{author}{Y.-P. Chang}, \bibinfo{author}{F.~Filsinger},
  \bibinfo{author}{B.~G. Sartakov}, \bibinfo{author}{J.~Küpper},
\newblock \bibinfo{title}{{CMIstark}: {P}ython package for the stark-effect
  calculation and symmetry classification of linear, symmetric and asymmetric
  top wavefunctions in dc electric fields},
\newblock \bibinfo{journal}{Comp.\ Phys.\ Comm.}  (\bibinfo{year}{2013}).
  \bibinfo{note}{{DOI}: 10.1016/j.cpc.2013.09.001}.
\bibitem[{Even et~al.(2000)Even, Jortner, Noy, Lavie, and
  Cossart-Magos}]{Even:JCP112:8068}
\bibinfo{author}{U.~Even}, \bibinfo{author}{J.~Jortner},
  \bibinfo{author}{D.~Noy}, \bibinfo{author}{N.~Lavie},
  \bibinfo{author}{N.~Cossart-Magos},
\newblock \bibinfo{title}{Cooling of large molecules below 1~{K} and {H}e
  clusters formation},
\newblock \bibinfo{journal}{J.\ Chem.\ Phys.} \bibinfo{volume}{112}
  (\bibinfo{year}{2000}) \bibinfo{pages}{8068--8071}.
\bibitem[{Fujimaki et~al.(1999)Fujimaki, Fujii, Ebata, and
  Mikami}]{Fujimaki:JCP110:4238}
\bibinfo{author}{E.~Fujimaki}, \bibinfo{author}{A.~Fujii},
  \bibinfo{author}{T.~Ebata}, \bibinfo{author}{N.~Mikami},
\newblock \bibinfo{title}{Autoionization-detected infrared spectroscopy of
  intramolecular hydrogen bonds in aromatic cations. {I}. principle and
  application to fluorophenol and methoxyphenol},
\newblock \bibinfo{journal}{J.\ Chem.\ Phys.} \bibinfo{volume}{110}
  (\bibinfo{year}{1999}) \bibinfo{pages}{4238--4247}.
\bibitem[{Manocha et~al.(1973)Manocha, Carlson, and
  Fateley}]{Manocha:JPC77:2094}
\bibinfo{author}{A.~S. Manocha}, \bibinfo{author}{G.~L. Carlson},
  \bibinfo{author}{W.~G. Fateley},
\newblock \bibinfo{title}{Barriers to internal rotation in some m-substituted
  phenols},
\newblock \bibinfo{journal}{J.\ Phys.\ Chem.} \bibinfo{volume}{77}
  (\bibinfo{year}{1973}) \bibinfo{pages}{2094--2098}.
\bibitem[{Holmegaard et~al.(2010)Holmegaard, Hansen, Kalh{\o}j, Kragh,
  Stapelfeldt, Filsinger, K\"upper, Meijer, Dimitrovski, Abu-samha, Martiny,
  and Madsen}]{Holmegaard:NatPhys6:428}
\bibinfo{author}{L.~Holmegaard}, \bibinfo{author}{J.~L. Hansen},
  \bibinfo{author}{L.~Kalh{\o}j}, \bibinfo{author}{S.~L. Kragh},
  \bibinfo{author}{H.~Stapelfeldt}, \bibinfo{author}{F.~Filsinger},
  \bibinfo{author}{J.~K\"upper}, \bibinfo{author}{G.~Meijer},
  \bibinfo{author}{D.~Dimitrovski}, \bibinfo{author}{M.~Abu-samha},
  \bibinfo{author}{C.~P.~J. Martiny}, \bibinfo{author}{L.~B. Madsen},
\newblock \bibinfo{title}{Photoelectron angular distributions from strong-field
  ionization of oriented molecules},
\newblock \bibinfo{journal}{Nat. Phys.} \bibinfo{volume}{6}
  (\bibinfo{year}{2010}) \bibinfo{pages}{428}.
\bibitem[{Filsinger et~al.(2011)Filsinger, Meijer, Stapelfeldt, Chapman, and
  K\"upper}]{Filsinger:PCCP13:2076}
\bibinfo{author}{F.~Filsinger}, \bibinfo{author}{G.~Meijer},
  \bibinfo{author}{H.~Stapelfeldt}, \bibinfo{author}{H.~Chapman},
  \bibinfo{author}{J.~K\"upper},
\newblock \bibinfo{title}{State- and conformer-selected beams of aligned and
  oriented molecules for ultrafast diffraction studies},
\newblock \bibinfo{journal}{Phys.\ Chem.\ Chem.\ Phys.} \bibinfo{volume}{13}
  (\bibinfo{year}{2011}) \bibinfo{pages}{2076--2087}.

\end{thebibliography}
\end{document}